\documentclass[aps,prl, amsmath,amssymb, reprint]{revtex4-2}
\usepackage{amsthm}
\usepackage{mathtools}
\usepackage{microtype}
\usepackage[colorlinks=true, citecolor=blue, urlcolor=blue]{hyperref}
\newcommand{\ket}[1]{| #1 \rangle}

\newcommand{\rank}{\operatorname{rank}}

\begin{document}
\title{Quantifying Nonstabilizerness of Quantum Codes by Removing the Inert Background}
\author{Yuan Liu}
\author{Ke-Mi Xu}
\email{xukemi@bit.edu.cn}
\affiliation{MIIT Key Laboratory of Complex-field Intelligent Exploration, School of Optics and Photonics, Beijing Institute of Technology, Beijing 100081, China}

\begin{abstract}
Nonstabilizerness (magic) is the quantum resource that, together with stabilizer operations, makes universal quantum computation possible. However, quantifying nonstabilizerness is difficult. The standard measure sums over exponentially many Pauli operators, and for quantum codes no quantitative theory has been available. A structural observation removes this obstacle: the Clifford sector carries no magic and can be removed by a Clifford transformation, leaving an order-three object whose magic is an exactly evaluable fourth moment. Removing this inert background yields closed forms for several code families: a formula for all Dicke states, reducing a 100-qubit case from $4^{100}$ terms to a short binomial sum; a bound for cubic-phase codes (twisted quantum doubles and non-Abelian topological order), saturated only by the $D_4$ code; and a cyclic/zero criterion for group multiplication states. For codeword-stabilized (CWS) codes, the companion paper \cite{Liu26arXiv} carries this reduction to its conclusion: the most magical codes are exactly the Sidon sets, and the nonstabilizerness bounds the code's transversal non-Clifford power. Together these results reduce the computation of nonstabilizerness for a broad class of codes to finite classical counting problems, and identify the maximally magical codes among them.
\end{abstract}

\maketitle

Nonstabilizerness, or magic, quantifies the departure of a quantum state from the stabilizer polytope. Together with stabilizer operations, it is the resource that enables universal quantum computation \cite{Bravyi05PRA, Howard17Mar, Chitambar19RMP, Liu22PRXQ}. A convenient measure of nonstabilizerness is the second-order stabilizer R\'enyi entropy (SRE) \cite{Leone22Feb, Haug23PRXQ}
\begin{equation}
\label{eq:M2}
  M_2(\ket{\psi})=-\log_2\Big(2^{-n}\sum_{P\in\mathcal P_n}\langle\psi|P|\psi\rangle^4\Big),
\end{equation}
where $\mathcal P_n$ is the $n$-qubit Pauli group. The global phases of $P$ are suppressed, since they do not enter the fourth moment. For a generic state the sum runs over all $4^n$ Pauli operators, so a direct evaluation is impractical already for a few tens of qubits \cite{Xiao26Aug, Sierant26Quantum}.

Quantum codes stores logical information in their code states, which are protected by fault-tolerant devices. Quantifying the nonstabilizerness of quantum codes (i.e., nonstabilizerness of the code states) is thus of great importance. However, prior works provide exact values only in special cases, and for whole code families only qualitative facts. Closed forms are known for the $W$ state \cite{Odavic23SciPost, Catalano24arXiv} , for fixed-excitation Dicke \cite{Liu25TIMPS} and hypergraph states \cite{Chen24Quantum, Kagamihara26arXiv}, and the permutation-invariant machinery of \cite{Passarelli24PRA} evaluates symmetric states efficiently in a Dicke basis, but these do not extend to full families or to large codes. Twisted quantum doubles furnish non-Pauli topological stabilizer codes \cite{Fuente21Quantum,Ellison22PRXQ}, whose SRE encodes fusion rules \cite{Hoshino26Feb} and whose non-Abelian orders carry extensive, long-ranged magic \cite{Kobayashi26Jun, Zhang26arXiv, Korbany25Oct, Wei26arXiv}, but no exact values have been reported. Exact magic for these code families has therefore remained out of reach.

Efficient quantification requires simplifying the sum in Eq.~\eqref{eq:M2}. A Clifford unitary conjugates the Pauli group into itself and leaves $M_2$ invariant, so any part of a state generated by Clifford operations is invisible to $M_2$ and removable. For a wide range of states this inert part is exactly the quadratic (order-two), Clifford-implementable sector, and removing it isolates an order-three (cubic) object that carries all the magic. In the language of the Pauli and Clifford hierarchy \cite{Bravyi13Apr}, the Pauli group is a central extension of $\mathbb{F}_2^{2n}$ classified by the symplectic form, the Clifford layer is the quadratic sector, and magic is the order-three obstruction to quadraticity. 

Removing the inert background thus reduces the nonstabilizerness of each family to a finite, exactly evaluable quantity. The reduction is carried out by two fourth-moment dictionaries: on the \emph{indicator side} a code state reduces to a subset state whose magic is the additive energy of its support, and on the \emph{phase side} to a phase state whose magic is the rank data of a cubic Hessian. Applied to four code families, the dictionaries yield several results. For Codeword-stabilized (CWS) codes the reduction produces the companion dictionary \cite{Liu26arXiv} and makes nonstabilizerness of Kerdock codes \cite{Kerdock1972182, Hammons94TIT} exactly computable. For permutation-invariant codes we obtain a closed form for all Dicke states, valid for every excitation number $k$ and block length $n$, so that the magic of a $100$-qubit Dicke state reduces from a sum over $4^{100}$ terms to a short binomial sum. For cubic-phase codes, the decoupled form of twisted quantum doubles and non-Abelian topological order, we obtain the exact magic together with a lower bound saturated only by the determinant on $\mathbb{F}_2^3$ (the $D_4$ code). For group multiplication states the magic vanishes if and only if the group law is cyclic, and the minimal non-cyclic case carries exactly the magic of the Controlled-Controlled-Z (CCZ) state. Our work thus provides a general and efficient framework to quantify nonstabilizerness of quantum codes.

\textit{Decoupling the Quadratic Background---}
The decoupling rests on a degree filtration of the Pauli and Clifford hierarchy. We work on $n$ qubits and write $\mathbb{F}_2^n$ for the set of $n$-bit strings $x=(x_1,\dots,x_n)$ with $x_i\in\{0,1\}$. This is also the label space of the Pauli group. A Pauli operator is written as (up to a phase) $X^{u}Z^{z}=\prod_{i=1}^n X_i^{u_i}Z_i^{z_i}$, $u,z\in\mathbb{F}_2^n$, with $X\ket b=\ket{b\oplus1}$ and $Z\ket b=(-1)^b\ket b$. The commutation rule $X^{u}Z^{z}\,X^{u'}Z^{z'}=(-1)^{u\cdot z'+u'\cdot z}\,X^{u'}Z^{z'}\,X^{u}Z^{z}$ exhibits $\mathcal P_n$ as a central extension of $\mathbb{F}_2^{2n}$ by $\mathbb{Z}_2$, with extension class the symplectic form $\omega((u,z),(u',z'))=u\cdot z'+u'\cdot z$. This is the Heisenberg group. A Clifford unitary is one that conjugates $\mathcal P_n$ into itself; on labels it acts by a symplectic transformation $\mathrm{Sp}(2n,\mathbb{F}_2)$, up to a phase that is itself quadratic in the labels. In this sense ``Clifford'' and ``quadratic'' are synonymous, i.e., they define the order-two sector.

The order-one and order-two content of a state is read off from its phases. A stabilizer state in the computational basis has the form
\begin{equation}
  \ket{\psi}\propto\sum_{x\in\mathbb{F}_2^n}(-1)^{q(x)}\ket x,
\end{equation}
with $q$ a quadratic Boolean function, $q(x)=\sum_i a_i x_i+\sum_{i<j}b_{ij}x_ix_j$. The canonical example is the graph state of a graph $G$ (with the set of edges denoted by $E(G)$) \cite{Raussendorf01May, Briegel01Jan, Hein04PRA}, whose phase is $q_G(x)=\sum_{\{i,j\}\in E(G)}x_ix_j$; every quadratic phase is a graph state up to single-qubit Cliffords \cite{Dehaene03PRA}. The quadratic part of a phase is invisible to $M_2$, since the diagonal unitary $\prod_iZ_i^{a_i}\prod_{i<j}\mathrm{CZ}_{ij}^{b_{ij}}$ that implements $(-1)^{q(x)}$ is Clifford, and $M_2$ is Clifford-invariant \cite{Leone22Feb}. Hence the magic of $2^{-n/2}\sum_x(-1)^{f(x)}\ket x$ is determined solely by the part of $f$ of degree at least three.

To isolate the high-degree part we introduce the discrete (Boolean) derivatives. For $f:\mathbb{F}_2^n\to\mathbb{F}_2$ the discrete derivative in direction $u$ is defined by $\Delta_u f(x)=f(x)+f(x+u)$. It is nilpotent ($\Delta_u^2=0$), whereas derivatives in distinct directions commute but do not cancel ( $\Delta_u\Delta_v\neq0$ for $u\neq v$). Iterating them therefore gives the generally nonzero second and third derivatives
\begin{equation}
\begin{aligned}
  \Delta_u\Delta_v f(x)&=f(x)+f(x{+}u)+f(x{+}v)+f(x{+}u{+}v),\\
  \Delta_u\Delta_v\Delta_w f&=\sum_{S\subseteq\{u,v,w\}}f\Big(\sum_{s\in S}s\Big),
\end{aligned}
\end{equation}
the former being the ``parallelogram curvature'' of $f$. Nilpotency acts only on repeated directions, so $\Delta_u\Delta_v\Delta_w f$ is an alternating trilinear form in $u,v,w$, vanishing whenever two of them coincide (hence it is a three-cocycle in the language of group cohomology). The order-two sector is invisible to $M_2$, and what survives is the three-cocycle $\Delta_u\Delta_v\Delta_w f$.

\begin{figure}
  \centering
  \includegraphics[width=8cm]{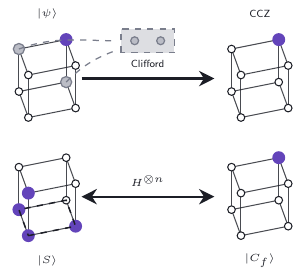}
  \caption{Decoupling the quadratic background. (a) The phase $(-1)^{f(x)}$ of a code state $\ket\psi=\sum_{x\in\mathbb F_2^3}(-1)^{f(x)}\ket x$ separates into a quadratic part (gray vertices) and a cubic core (violet). A Clifford unitary removes the quadratic background, leaving the cubic core. For $n=3$ the unique cubic is the determinant $f=x_1x_2x_3$, the CCZ state. (b) The surviving cubic object appears in two complementary forms: the subset state $\ket S$ (left; the Sidon set $S=\{000,100,010,001\}$, whose broken parallelogram marks the absence of additive structure) and the phase state $\ket{C_f}$ (right, the CCZ state).}
\label{fig:peeling}
\end{figure}

We therefore ask: after a Clifford transformation that removes the quadratic background, what does the code state become? For all the codes we consider, the answer is one of two canonical forms (Fig.~\ref{fig:peeling}). On the indicator side the state is a subset state
\begin{equation}
  \ket S=|S|^{-1/2}\sum_{x\in S}\ket x,
\end{equation}
whose magic is governed by the nonlinearity of the support $S$. On the phase side the state is a phase state
\begin{equation}
  \ket{C_f}=2^{-n/2}\sum_{x\in\mathbb{F}_2^n}(-1)^{f(x)}\ket x,
\end{equation}
whose magic is governed by the cubic part of $f$. Each canonical form admits an exact fourth-moment dictionary. We derive the two dictionaries in the next section. The applications then require identifying, for each code family, which canonical form its code state reduces to.

\textit{Two Fourth-moment Dictionaries---}
The derivation depends on the fourth-moment form of Parseval's identity. For any real-valued function $a:\mathbb{F}_2^n\to\mathbb{R}$, we have
\begin{equation}
\label{eq:parseval}
  \sum_{z\in\mathbb{F}_2^n}\Big|\sum_{x\in\mathbb{F}_2^n}a(x)(-1)^{z\cdot x}\Big|^4=2^n\sum_{v\in\mathbb{F}_2^n}\Big|\sum_{x\in\mathbb{F}_2^n}a(x)a(x\oplus v)\Big|^2 .
\end{equation}

For a set $S\subset\mathbb{F}_2^n$ with state $\ket S=|S|^{-1/2}\sum_{x\in S}\ket x$ and a Pauli operator $P=X^uZ^z$, the expectation value is
\begin{equation}
  \langle S|X^uZ^z|S\rangle=\frac1{|S|}\sum_{x\in S\cap(S\oplus u)}(-1)^{z\cdot x}.
\end{equation}
Fixing $u$ and putting $S_u=S\cap(S\oplus u)$, applying \eqref{eq:parseval} to $a=\mathbf1_{S_u}$ gives $\sum_z|\langle S|X^uZ^z|S\rangle|^4=|S|^{-4}2^n E(S_u)$, where
\begin{equation}
  E(S)=\sum_{y\in\mathbb{F}_2^n}|S\cap(S\oplus y)|^2
\end{equation}
is the additive energy of $S$ \cite{Tao_Vu_2006}. Summing over $u$ and inserting into \eqref{eq:M2} yields the indicator-side dictionary,
\begin{equation}
\label{eq:indicator}
\begin{aligned}
  M_2(\ket S)&=4\log_2|S|-\log_2 E^{(1)}(S),\\
  E^{(1)}(S)&=\sum_{x\in\mathbb{F}_2^n}E(S\cap(S\oplus x)).
\end{aligned}
\end{equation}
This is a fourth moment of the difference structure: a parallelogram $\{x,x{+}u,x{+}v,x{+}u{+}v\}$ contributes to $E^{(1)}$ exactly when all four corners lie in $S$. A linear code $S$ has $E^{(1)}=|S|^4$ and hence $M_2=0$, as it must; magic arises only from the \emph{nonlinearity} of $S$. This dictionary is the tool behind the CWS analysis of the companion paper \cite{Liu26arXiv}, and it applies verbatim to any code whose code state is Clifford-equivalent to a subset state.

The phase side is the Fourier-dual counterpart of the indicator side. Instead of a sparse $\{0,1\}$ indicator on a subset $S$, the relevant function takes both signs, $a=\pm1$, on all of $\mathbb{F}_2^n$, and the identical Parseval identity \eqref{eq:parseval} supplies its dictionary. For $P=X^uZ^z$ we have
\begin{equation}
\begin{aligned}
  \langle C_f|X^uZ^z|C_f\rangle&=2^{-n}\sum_x(-1)^{f(x)+f(x+u)+z\cdot x}\\
  &=2^{-n}\sum_x a_u(x)(-1)^{z\cdot x},
\end{aligned}
\end{equation}
with $a_u(x)=(-1)^{\Delta_uf(x)}$ real. Applying Eq.~\eqref{eq:parseval} and $a_u(x)a_u(x+v)=(-1)^{\Delta_u\Delta_v f(x)}$ gives
\begin{equation}
\label{eq:phase_dict}
  M_2(\ket{C_f})=n-\log_2\Big(2^{-3n}\sum_{u,v\in\mathbb{F}_2^n}\Big|\sum_x(-1)^{\Delta_u\Delta_v f(x)}\Big|^2\Big),
\end{equation}
a fourth moment of the second derivative of $f$. 

For a cubic $f$, $\Delta_u\Delta_v\Delta_w f$ does not depend on the base point $x$, and $\Delta_u\Delta_v f(x)$ is affine in $x$. An affine $g$ satisfies $g(x)=g(0)\oplus\Delta_x g(0)$, so for $g=\Delta_u\Delta_v f$, we have $\Delta_u\Delta_v f(x)=\Delta_u\Delta_v\Delta_x f\oplus c(u,v)$, $c(u,v):=\Delta_u\Delta_v f(0)$. The $x$-dependent term is the three-cocycle $B(u,v,x):=\Delta_u\Delta_v\Delta_x f$, and the constant $c(u,v)$ is irrelevant as it only flips the overall sign of the Fourier coefficient. The sum over $x$ then selects $B(u,v,\cdot)\equiv0$, and
\begin{equation}
\label{eq:cubic_dict}
\begin{aligned}
  M_2(\ket{C_f})&=2n-\log_2 N(B),\\
  N(B)&=\#\big\{(u,v)\in\mathbb{F}_2^{2n} \mid B(u,v,\cdot)\equiv0\big\}.
\end{aligned}
\end{equation}
Eq.~\eqref{eq:cubic_dict} is the degree-three instance of the hypergraph-state SRE formula of Ref.~\cite{Chen24Quantum}, in the rank form given in Ref.~\cite{Kagamihara26arXiv}; within the two-dictionary framework it is recovered from the same Parseval identity as the indicator side, and expressed through the trilinear form $B$. The magic of a cubic code is therefore determined entirely by the trilinear form $B$. Geometrically, $N(B)$ counts singular directions of the Hessian. For each $u$ the slice $M_u(v,w)=B(u,v,w)$ is a bilinear form on $\mathbb{F}_2^n$, and $B(u,v,\cdot)\equiv0$ means $v\in\ker M_u$, so $N(B)=\sum_u2^{n-\rank M_u}$.

Each slice is \emph{alternating}, i.e., $M_u(v,v)=B(u,v,v)=B(v,v,u)$, and $B(v,v,\cdot)=\Delta_v^2f(\cdot)$ vanishes identically on $\mathbb{F}_2$. Consequently $M_u$ is alternating and $\rank M_u$ is even for every $u$. The direction $u=0$ contributes $2^n$; every nonzero direction contributes at least $1$ (even $n$) or $2$ (odd $n$, where the maximal even rank is $n-1$). Hence
\begin{equation}
  N(B)\ge\begin{cases}2^{n+1}-1,& n\ \text{even},\\ 3\cdot2^n-2,& n\ \text{odd}.\end{cases}
\end{equation}
The bound is attainable only if every $M_u$ with $u\neq0$ is nonsingular. For $n=3$ the only cubic form (up to scaling) is the determinant $f=x_1x_2x_3$, for which $M_u(v,w)=\det(u,v,w)$ is a nonvanishing alternating form on $\mathbb{F}_2^3$, of rank $2=n-1$ whenever $u\neq0$. Thus $N(\det)=2^3+7\cdot2^{3-2}=22$, and the bound is saturated:
\begin{equation}
  M_2(\mathrm{CCZ})=6-\log_2 22,
\end{equation}
which coincides with the value obtained for the three-qubit CCZ (hypergraph) state in Ref.~\cite{Chen24Quantum}. The maximally magical cubic form is thus unique and three-dimensional. On the phase side it is the CCZ state, while on the indicator side the corresponding Sidon set \cite{Sidon1932, Czerwinski24AdvMathCommun} $S=\{000,100,010,001\}$ realizes the same extremal value $M_2=6-\log_222$. In general the bound need not be tight. We have verified exhaustively for $n=4,5$ that the bound is not attainable \footnote{Uniqueness in dimension three and non-attainability for $n\ge4$ are classical in the theory of trilinear forms. Over $\mathbb F_2$ an alternating trilinear form has a singular line unless $n=3$, where the unique exception is the determinant. This is the theorem of Draisma and Shaw~\cite{DraismaShaw10}, which applies to finite fields via the Chevalley--Warning theorem and builds on the classical classification of vector cross products in dimensions three and seven~\cite{Hurwitz1898, Eckmann1942, BrownGray67, Massey83}. To our knowledge this is the first translation of that classification into the extremization of the stabilizer R\'enyi entropy and the explicit bound above.}.

\textit{Codeword-stabilized Codes---}
CWS codes were analyzed in detail in the companion paper \cite{Liu26arXiv}, so we only recall the point of contact. A CWS code is a graph state $\ket G$ dressed by a classical code $C\subset\mathbb{F}_2^n$ \cite{Cross09TIT, Chuang09JMP}, with code state $\ket C=|C|^{-1/2}\sum_{w\in C}Z^w\ket G$. The graph is the quadratic background. Since $HZH=X$, a global Hadamard sends each codeword to $X^w H^{\otimes n}\ket G$, and by writing $\ket G=\prod_e\mathrm{CZ}_e\ket{+}^{\otimes n}$ one has $H^{\otimes n}\ket G=\tilde U_G\ket0^{\otimes n}$ with $\tilde U_G$ a Clifford circuit. Thus the code state reduces to a subset state up to Clifford,
\begin{equation}
  M_2(\text{CWS code}) =4\log_2|C|-\log_2 E^{(1)}(C).
\end{equation}
The graph is gauge, and the magic is the additive-combinatorial nonlinearity of the classical label code. This is the origin of the Sidon and Kerdock structure in the companion paper.

\textit{Permutation-invariant Codes and Dicke States---}
A permutation-invariant code has a code space closed under arbitrary qubit permutations \cite{Ouyang14PRA}. Its canonical basis states are the Dicke states \cite{Dicke54PR}, and a single Dicke state is the subset state of the weight-$k$ layer,
\begin{equation}
\ket{D^k_n}=\binom nk^{-1/2}\sum_{|x|=k}\ket x .
\end{equation}
Permutation symmetry has already eliminated the geometric background, so the decoupling reduces to evaluating the additive energy of a weight layer, and Eq.~\eqref{eq:indicator} applies with $|C_k|=\binom nk$. The energy is tractable because it factorizes. For a vector $x_{2j}$ of weight $2j$, the slice $C_k\cap(C_k\oplus x_{2j})$ consists of those $x$ with $|x|=k$ and $|x\oplus x_{2j}|=k$; since $|x\oplus x_{2j}|=|x|+2j-2|x\cap x_{2j}|$, the latter means $x$ has exactly $j$ of its ones inside $\mathrm{supp}(x_{2j})$. The $n$-cube thus splits into the $2j$ coordinates of $\mathrm{supp}(x_{2j})$ and the remaining $n-2j$ coordinates, and $C_k\cap(C_k\oplus x_{2j})$ is the product of a weight-$j$ layer of $\mathbb{F}_2^{2j}$ with a weight-$(k-j)$ layer of $\mathbb{F}_2^{n-2j}$. The additive energy factorizes over a product, $E(A\times B)=E(A)E(B)$, and the energy of a weight-$r$ layer of the $m$-cube is elementary: for $|y|=2t$ one must place $t$ of its ones in $\mathrm{supp}(y)$ and $r-t$ in the complement. Let $L(r,m):=E(\text{weight-}r\ \text{layer of}\ \mathbb{F}_2^m)$, then
\begin{equation}
  L(r,m)=\sum_{t=0}^{r}\binom m{2t}\binom{2t}{t}^2\binom{m-2t}{r-t}^2.
\end{equation}
Putting the pieces together, we obtain
\begin{equation}
  E^{(1)}(C_k)=\sum_{j=0}^{k}\binom n{2j}\,L(j,2j)\,L(k-j,n-2j),
\end{equation}
and therefore, for all $n,k$,
\begin{equation}
\label{eq:dicke}
\begin{aligned}
    &M_2(\ket{D^k_n})\\
    ={}&4\log_2\binom nk -\log_2\Big[\sum_{j=0}^{k}\binom n{2j}L(j,2j)L(k-j,n-2j)\Big].
\end{aligned}
\end{equation}
We verify the formula in two cases. For $k=1$ (the $W$ state) one has $L(1,n)=3n^2-2n$ and $L(1,2)=8$, so $E^{(1)}(C_1)=L(1,n)+\binom n2\cdot8=7n^2-6n$ and $M_2(W_n)=4\log_2n-\log_2(7n^2-6n)=3\log_2n-\log_2(7n-6)$, the known result \cite{Odavic23SciPost, Catalano24arXiv}; for $k=2$ one recovers $E^{(1)}(C_2)=\tfrac14n(n-1)(91n^2-427n+492)$ \cite{Liu26arXiv, Liu25TIMPS}, and the case $k=3$ likewise appears in Ref.~\cite{Liu25TIMPS}. These cases exhaust the previously known results, yet Eq.~\eqref{eq:dicke} is exact for all $n,k$ and extends them to arbitrary excitation number. For example, the values
\begin{equation}
  M_2(D^{10}_{100})=44.53,\qquad M_2(D^{20}_{200})=85.73,
\end{equation}
whose naive evaluation would require summing $4^{100}$ and $4^{200}$ Pauli operators, are each a short binomial sum now.

\textit{Cubic-phase Codes and non-Abelian Topological Order---}
The phase dictionary applies directly to cohomological code families. A twisted quantum double $D^\omega(G)$ is defined by a three-cocycle $\omega\in H^3(G,\mathrm{U}(1))$. After the untwisted (stabilizer) background is trivialized, its code state is a phase state with cubic $f$, the phase of the cocycle \cite{Kobayashi26Jun,Ellison22PRXQ}. The magic is then a statement about the trilinear form $B$ alone. The minimal instance is the twisted $\mathbb{Z}_2^3$ double, for which $H^3(\mathbb{Z}_2^3,\mathrm{U}(1))=\mathbb{Z}_2$ and the nontrivial cocycle is the triple cup product $\omega=a\cup b\cup c$. This model is equivalent to the $D_4$ topological order \cite{Kobayashi26Jun}, and its decoupled state is the determinant/CCZ state, with magic $M_2=6-\log_2 22$. More generally, non-Abelian topological order carries intrinsic magic. A two-dimensional order is realizable by Pauli stabilizers iff its anyons are Abelian, and non-Abelian string-net ground states have extensive, long-range magic that no finite-depth local unitary can remove \cite{Ellison22PRXQ,Zhang26arXiv}.

As a benchmark family of exactly solvable cubic codes we single out $f=x_0 Q(\mathbf y)$, the product of a linear coordinate $x_0$ with a quadratic form $Q$ on the remaining $m=n-1$ coordinates whose polar form $B_Q$ has maximal rank, i.e., $\rank B_Q=m$ for even $m$ and $\rank B_Q=m-1$ for odd $m$ (note that $B$ depends on $Q$ only through $B_Q$). By the Leibniz rule for discrete derivatives, we have
\begin{equation}
  B(u,v,w)=u_0 B_Q(v_y,w_y)+v_0 B_Q(w_y,u_y)+w_0 B_Q(u_y,v_y),
\end{equation}
where $u=(u_0,u_y), v= (v_0, v_y)$, and $w=(w_0, w_y)$. The condition $B(u,v,\cdot)\equiv0$ splits into $B_Q(u_y,v_y)=0$ and $u_0B_Q(v_y,\cdot)+v_0B_Q(\cdot,u_y)\equiv0$, and counting the pairs $(u,v)$ (distinguishing $u_y=0$, $u_y$ in the radical of $B_Q$, and $u_y$ generic) gives
\begin{equation}
\label{eq:x0Q}
\begin{aligned}
  N(B)&=\begin{cases}2^{m+2}+2^{2m-1}-2^{m-1},& m\ \text{even},\\
  2^{m+3}+2^{2m-1}-2^{m},& m\ \text{odd},\end{cases}
  \\
  M_2&=2(m{+}1)-\log_2 N(B).
\end{aligned}
\end{equation}
For $m=2,4,6,8$ this gives $N=22,184,2272,33664$, i.e., $M_2\approx 1.541,2.476,2.850,2.961$. The family is the phase-side counterpart of the Kerdock construction---a linear coordinate times a symplectic form, mirroring how the Kerdock set is built from skew (symplectic) matrices \cite{Calderbank97ProcLondonMathSoc}. Moreover, it is easy to directly check that for $n\le5$ it is the exact maximizer of nonstabilizerness.

\begin{table*}
\caption{Summary of code families, their inert backgrounds, and the corresponding magic carriers.}
\begin{tabular}{lll}
\hline
code family & inert background  & magic carrier \\
\hline
CWS codes & graph state (quadratic form) & classical code $C$ (additive energy) \\
permutation-invariant & symmetric (empty/complete) graph & weight profile (layer energy) \\
cubic / cohomological & quadratic phase (Clifford sector) & three-cocycle $\Delta_u\Delta_v\Delta_w f$ (rank data $N(B)$) \\
group algebra ($D(G)$) & cyclic group law (Abelian sector) & non-cyclic / non-Abelian law (cubic defect) \\
\hline
\end{tabular}
\label{tab:Summary}
\end{table*}

\textit{Non-Abelian Group States---}
As a further instance we isolate the group-theoretic core of a quantum double $D(G)$ \cite{Kitaev03AnnPhys}, the \emph{multiplication state}
\begin{equation}
  \ket{\psi_G}=|G|^{-1}\sum_{g,h\in G}\ket g\ket h\ket{gh}
\end{equation}
on three qudits of dimension $|G|$, i.e., the image of $\ket{+}\ket{+}\ket0$ under the ``controlled group multiplication'' gate $U_{\mathrm{mult}}\ket{g,h,0}=\ket{g,h,gh}$. The qudit Pauli frame is fixed by the cyclic shift $X\ket j=\ket{j+1 \mod d}$, $d=|G|$, and all Pauli operators below refer to this cyclic frame. Relative to this frame the state carries magic precisely when the group law is not cyclic, while for a cyclic group the state is manifestly stabilizer. With $gh=(g+h) \mod d$,
$\ket{\psi_G}=d^{-1}\sum_{g,h}\ket{g,h,g+h}$ is fixed by the $d^3$ commuting generalized Paulis
\begin{equation}
  \{X^a\otimes X^b\otimes X^{a+b}\}_{a,b\in\mathbb{Z}_d}\ \cup\ \{(Z\otimes Z\otimes Z^{-1})^t\}_{t\in\mathbb{Z}_d}.
\end{equation}
The first family merely permutes the terms of the uniform superposition and the second has eigenvalue $\omega^{t(g+h-(g+h))}=1$ on the support. The two families commute since $\omega^{at+bt-(a+b)t}=1$. Conversely, if the group law is not cyclic, then for some $a,b,c$ the diagonal expectation $\langle\psi_G|Z^aZ^bZ^c|\psi_G\rangle=d^{-2}\sum_{g,h}\omega^{ag+bh+c(gh)}$ has magnitude strictly between $0$ and $1$, because $(g,h)\mapsto\omega^{c(gh)}$ extends to a character of $G\times G$ only for a cyclic law. Such a Pauli expectation is forbidden for a stabilizer state, so the state is non-stabilizer. Hence $M_2(\ket{\psi_G})=0$ if and only if the group law is cyclic.

The minimal non-cyclic group is the Klein four-group, and it exposes the bridge to the cubic picture. For $G=\mathbb{Z}_2\times\mathbb{Z}_2$ labeled by $0\mapsto e,\ 1\mapsto a,\ 2\mapsto b,\ 3\mapsto ab$, the Klein law is $\mathbb{F}_2^2$ addition. In the $\mathbb{F}_2^2$ (two-qubit) frame the state factors as $\ket{\psi_G}=\ket{\psi_\mathrm{add}}^{\otimes2}$, with $\ket{\psi_\mathrm{add}}=2^{-1}\sum_{a,b}\ket a\ket b\ket{a+b}$ a stabilizer (parity) state. The non-trivial content appears only in the cyclic $\mathbb{Z}_4$ frame, where the defect of $\mathbb{F}_2^2$ addition relative to cyclic addition is the triple cup product. The fourth-moment sum evaluates to $22$, the dependent-pair count of the determinant on $\mathbb{F}_2^3$, so that
\begin{equation}
  M_2(\ket{\psi_{\mathbb{Z}_2\times\mathbb{Z}_2}})=6-\log_222,
\end{equation}
which is exactly the CCZ value. For larger non-Abelian groups the magic is non-vanishing, with a magnitude depending on the labeling. In a fixed natural labeling 
\footnote{Here a group of order $d$ is labeled by its elements $0,\dots,d-1$ in the order given below, relative to the cyclic qudit Pauli frame $X\lvert j\rangle=\lvert j+1\bmod d\rangle$, $Z\lvert j\rangle=\omega^{j}\lvert j\rangle$ with $\omega=\mathrm{e}^{2\pi \mathrm{i}/d}$. For $S_3=\langle r,s\mid r^3=s^2=1,\ srs=r^{-1}\rangle$ the element $r^k s^m$ carries the label $k+3m$; for the dihedral group $D_4=\langle r,s\mid r^4=s^2=1,\ srs=r^{-1}\rangle$ of order $8$ (not the $D_4$ topological-order code of the preceding section) it carries $k+4m$; and for $Q_8=\{\pm1,\pm i,\pm j,\pm k\}$ the order is $1,-1,i,-i,j,-j,k,-k$.}
one finds $M_2(S_3)=4.38$, $M_2(D_4)=4.39$, $M_2(Q_8)=6.13$, while $M_2(\mathbb{Z}_3)=M_2(\mathbb{Z}_4)=M_2(\mathbb{Z}_5)=M_2(\mathbb{Z}_6)=0$. The magnitude does not reduce to a single elementary group invariant ($D_4$ and $Q_8$ share order and number of conjugacy classes yet differ), reflecting that the magic is a genuine non-Abelian additive energy of the multiplication map.

Table~\ref{tab:Summary} summarizes the magic structure of different code families. The essential point is that the inert background is Clifford and removable, while the magic is cubic and countable. This converts an exponential Pauli sum into finite classical counting, gives exact formulae for entire code families, identifies their maximally magical members, and supplies a unified quantitative framework for nonstabilizerness in fault-tolerant quantum codes.

\medskip
\begin{acknowledgments}
\textit{Acknowledgments---}This work is supported by the Fundamental and Interdisciplinary Disciplines Breakthrough Plan of the Ministry of Education of China (grant Nos.~JYB2025XDXM115 and JYB2025XDXM201) and the Beijing Science and Technology Planning Project (grant No.~Z25110100040000).

\textit{Data Availability---}No data were created or analyzed in this study.
\end{acknowledgments}

\end{document}